\documentclass[aps,prb,reprint,amssymb,amsfonts]{revtex4-1} 
\usepackage{mathptmx} 
\usepackage{bm}
\usepackage{graphicx}
\usepackage{hyperref}
\hypersetup{
pdfborder={0 0 0}, 
colorlinks=true, 
linkcolor=blue,
citecolor=blue,
urlcolor=blue
}

\begin{document}

\title{Steady-state spectrum of Kelvin waves on a quantized vortex at finite temperatures}

\author{R. H\"anninen}
\affiliation{Low Temperature Laboratory, Aalto University, PO BOX 15100, 00076 AALTO, Finland}

\date{\today}

\begin{abstract}
We determine numerically the steady-state spectrum for Kelvin waves on a superfluid $^4$He 
vortex that is driven by shaking its end points and damped by mutual friction. 
The relaxation rate towards a steady state is determined by the mutual friction. We have reached 
the zero temperature limit where the steady-state spectrum and the vortex length become independent 
of the temperature. Our drive with pointlike pinning sites produces a characteristic spectrum with 
$|w_m|\propto m^{-\eta}$, where $\eta \approx$ 1.88. The spatially sharply peaked drive and 
the fact that we are fixing the oscillation amplitude, not the power, causes that there exists 
no high $k$ cutoff from mutual friction even when mutual friction is large. This spectrum is generic
to the pinned boundary conditions used. Without pinning the spectrum is sensitive to the drive. The
spectrum even depends on the amplitude of the initial spectrum if vortex is allowed to relax on its 
own. Therefore, in real systems the Kelvin spectrum is not unique and depends on external conditions.   
\end{abstract}
%
\maketitle

\section{Introduction}\label{s.intro}

During the last few years a central issue in the research on quantum turbulence (QT) has been 
the free decay of a vortex tangle. An important part of the decay at low temperatures is expected 
to be due to the Kelvin-wave cascade that works on length scales smaller than the intervortex 
distance. Due to circulation quantization the classical Kolmogorov decay cannot work at these scales. 

It is expected that at the lowest temperatures, where the dissipation due to mutual friction (m.f.) 
is negligible, the Kelvin cascade pushes the energy down to length scales of order the vortex 
core where it can be dissipated by phonons. However, the Kelvin-wave cascade works also at 
finite temperatures with nonzero mutual friction but now there should exist a cut-off introduced
by the m.f. Here we consider nonzero temperatures, since by doing so we avoid the difficulty to 
model the dissipation at the smallest scales. 

Kelvin waves (KWs) are helical distortions on a vortex \cite{Kelvin}.
Coupling between different Kelvin modes results in a cascade from large to smaller scales. 
Currently there are two main competing theories that predict a slightly different exponent for 
the Kelvin-wave spectrum \cite{KS2004prl,LN2010}. There are also numerical calculations 
\cite{KS2005prl,VinenPRL2003,KS2010prbsub,Baggaley2012}, but the correct spectrum is still under active 
debate \cite{Lvov2010JLTP,KS2010JLTPcomment,Lvov2010JLTPreply,KS2011prb,LLNR2010prb,Lvov2011prb,Sonin2012}. 
In the latest paper by Sonin\cite{Sonin2012} both theories are being criticized but the correct 
exponent for the Kelvin spectrum is argued to be given by Kozik's and Svistunov's model, provided
that the Kelvin amplitude is small enough. 

Here we calculate the Kelvin spectrum in a experimentally feasible situation where 
the Kelvin waves are exited by physically shaking the vortex. We are most interested in the 
zero temperature limit. Our results indicate that the spectrum depends only weakly on mutual 
friction. Therefore we expect that our results are also valid in the limit $T \rightarrow 0$. 

We also illustrate, by changing the drive, that the obtained spectrum is very generic, provided 
that the vortex end points are kept pinned. Without pinning the spectrum is much more sensitive to 
the drive. Also, if the vortex is allowed to relax on its own, the spectrum at later times 
depends on the amplitude of the initial spectrum.

\section{Method}

To obtain and accurately determine the steady state energy spectrum one should be able to 
simulate a large range of scales. Even nowadays this is quite difficult when we have a large 
vortex tangle with hundreds of vortices. To obtain a steady state, one should additionally prepare a drive, 
that typically works on large scales, and dissipation that is strongest at smallest scales. 
Since we are interested in the Kelvin cascade, we simplify the situation and consider only a 
single vortex in a configuration similar to that in Refs.~[\onlinecite{VinenPRL2003,KS2005prl,KS2010prbsub}]. 
Instead of simply removing the smallest scales, we damp the smallest scales by including
mutual friction. To drive the Kelvin waves we force the end points of the vortex to move
sinusoidally perpendicular to the direction of the unperturbed straight vortex, i.e. 
$x$ = $x_0\sin(\omega t)$, $y=0$, for points $z=0,L_z$. 
We still assume periodic boundary conditions (b.c.) along the $z$-direction. 
This corresponds to a periodic grid of pinning points moving coherently, which is somewhat
reminiscent of an oscillating grid where a vortex extends over several grid holes. The 
pinning points are considered to be small. The frequency of oscillation, $\omega$, is adjusted to be 
close to the resonance frequency of the Kelvin wave, which for wave vector $k_m$ is given by \cite{Kelvin}
\begin{equation}\label{e.kelvinfreq}
\omega_m = \frac{\kappa k_m^2}{4\pi}\lbrack \ln(2/k_ma) - \gamma \rbrack\,.
\end{equation}
Here $\kappa = h/m_4 \approx 0.0997$ mm$^2$/s is the circulation quantum,  $a \approx$ 
10$^{-7}$ mm is the vortex core size, and $\gamma = 0.57721\ldots$ is the Euler constant.
The above equation is valid for small amplitude Kelvin waves in the long wavelength limit 
($k_m a \ll 1$).
Due to periodic boundary conditions the wave vector is discrete and given by $k_m = 2\pi{m}/L_z$, 
where $m \in \mathbb{Z}$. 

We use the vortex filament model where the vortex is considered to be thin \cite{schwarz85}.  
By adopting the scale separation scheme developed by Kozik and Svistunov \cite{KS2005prl} we are able
to use much better resolution than in the earlier work by Vinen {\it et al.} \cite{VinenPRL2003}
One should note that even if the scale separation scheme reduces the amount of work done
(on average) per one time iteration from $N^2$ to $N\ln{N}$, where $N$ is the number of 
points on the vortex, the number of operations for a fixed time window scales like 
$N^3\ln{N}$ (without scale separation the work is proportional to $N^4$). This is 
because the numerical time step scales like $(L/N)^2$, where $L$ is the vortex length.
The scale separation routine has been criticized since the long and short scale dynamics 
are partly disconnected and therefore it could give incorrect spectrum \cite{Lvov2010JLTPreply}. 
We have verified that in our case the obtained spectra with and without the scale separation
scheme are identical when using a ``medium'' resolution ($N$=1024).

Our numerical scheme is standard practice in the vortex filament formulation except 
for the addition of the scale separation scheme. We use the standard 4th order Runge-Kutta 
method and adapt many of the discretization methods described in the Ph.D. Thesis 
of Aarts \cite{AartsThesis}. The superfluid velocity at point ${\bf s}$ is given
by the Biot-Savart equation:
\begin{eqnarray}\label{e.bs}
{\bf v}_{s}({\bf s},t) &=& \frac{\kappa}{4\pi}\hat{\bf s}'\times {\bf s}'' 
\ln\left(\frac{2\sqrt{l_{+}l_{-}}}{e^{1/2}a}\right) \\
&+&
\frac{\kappa}{4\pi}\int^{'}\frac{({\bf s}_1-{\bf s})\times d{\bf s}_1}
{\vert {\bf s}_1-{\bf s}\vert^3}\,. \nonumber
\end{eqnarray}
Here we have removed the singularity at ${\bf s}_1 = {\bf s}$ by dividing the 
integral into a local and non-local term as in Ref.~[\onlinecite{schwarz85}].
Vectors $\hat{\bf s}'$ and  ${\bf s}''$ in the local term are the tangent and the principal 
normal at ${\bf s}$ and the derivation is with respect to the arc length $\xi$.
Terms $l_{-}$ and $l_{+}$ are the lengths of two adjacent line elements connected 
to the point ${\bf s}$ after discretization. In the non-local term the integral 
is now over the remaining configuration. The numerical factor $e^{1/2}$ only 
takes into account the velocity distribution inside the core and is 
not essential here. 
At finite temperatures a vortex point moves at a velocity that is given by \cite{schwarz85}
\begin{equation}
{\bf v}_{\rm L}={\bf v}_{\rm s} +\alpha
\hat{\bf s}' \times ({\bf v}_{\rm n}-{\bf v}_{\rm s}) -\alpha'
\hat{\bf s}' \times [\hat{\bf s}' \times ({\bf v}_{\rm n}-{\bf v}_{\rm s})]\,,
\label{e.vl}
\end{equation}
where $\alpha$ and $\alpha'$ are the temperature dependent mutual friction coefficients. For 
simplicity, we set $\alpha' = 0$ which is appropriate for $^4$He at low temperatures.
We additionally assume that the normal fluid is at rest, ${\bf v}_n = 0$. The length (or period) of 
the unperturbed vortex is taken to be $L_z$ = 1 mm. The oscillation amplitude of the vortex 
end point is kept small enough to avoid reconnections and to keep the configuration such that 
we can describe it in terms of coordinates $x(z)$ and $y(z)$ as a sum of different Kelvin modes:
\begin{equation}\label{e.fourier}
x(z)+iy(z) = \sum_m w_m\exp({\rm i} 2\pi m z/L_z)\,.
\end{equation}
During simulations the Fourier 
components $w_m$ are determined by using a FFT. Since the vortex filament model does not 
automatically keep the points equally distributed along the $z$-axis we unify the grid before 
Fourier transformation by using a second order interpolation. The error here is negligible
since the difference from the uniform grid is small, provided that the grid unification is 
done frequently enough.

The power dissipated due to mutual friction can be estimated by calculating 
$P_{\rm mf} = \int({\bf v}_{\rm L}\cdot {\bf f}_{\rm mf})d\xi = 
-\int({\bf v}_{\rm L}\cdot {\bf f}_{\rm M})d\xi$, where ${\bf f}_{\rm mf}$ is force on a vortex
due to mutual friction and ${\bf f}_{\rm M}$ is the Magnus force, both given per unit
length (see Refs.~[\onlinecite{schwarz85,Donnelly}]). Assuming that ${\bf v}_n = 0$ the
power dissipated becomes:
\begin{equation}\label{e.Pmf}
P_{\rm mf} = -\alpha\rho_s\kappa \int|\hat{\bf s}'\times{\bf v}_s|^2d\xi\,.
\end{equation}

\begin{figure}[t]
\includegraphics[width=0.99\linewidth]{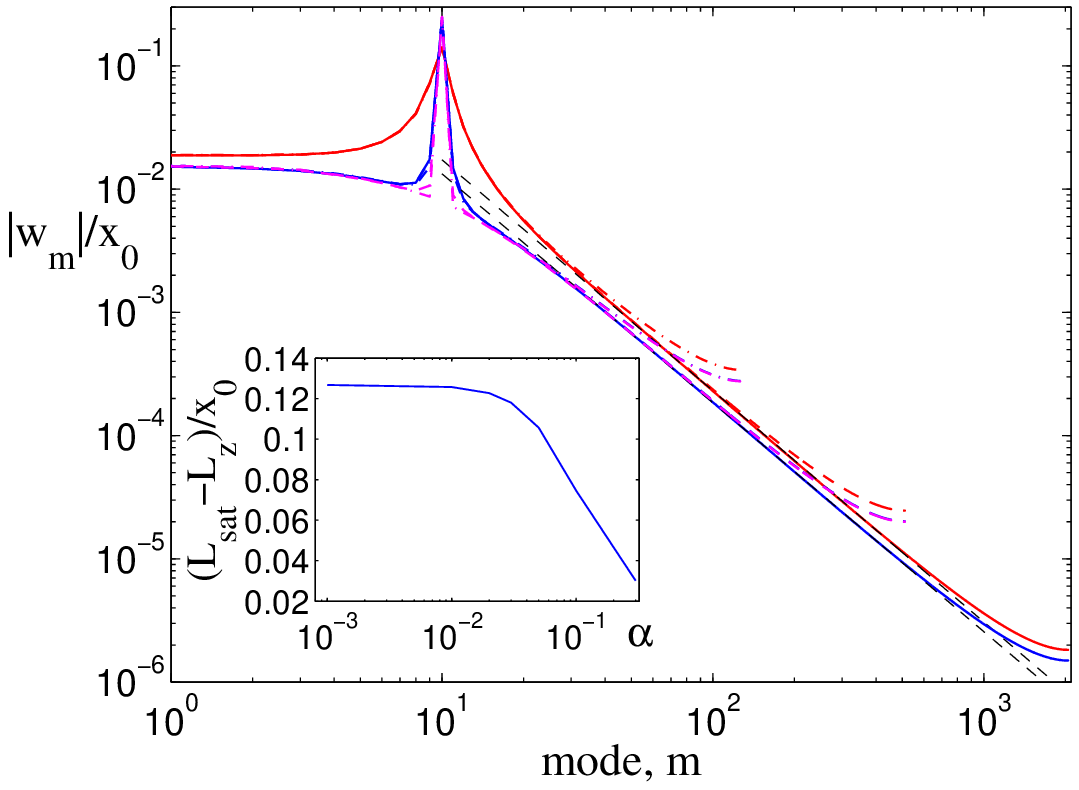}
\caption{(Color online) 
Steady-state Kelvin spectrum, averaged over the oscillation period of the drive. The drive 
is adjusted to be near mode $m$ = 10 ($\omega = \omega_m$) with amplitude $x_{0}=0.0005$ mm.
The red curves are for $\alpha$ = 0.1, blue for $\alpha$ = 0.01, 
and magenta (in many places overlapping with blue) for $\alpha$ = 0.001. 
The simulations have been done using 256 (dash-dotted), 1024 (dashed), and 4096 (solid) 
points, which correspond to different cutoffs. The black dashed lines correspond to fits 
for $\alpha$ = 0.1 and $\alpha$ = 0.01. The resulting slopes for $|w_m|\propto m^{-\eta}$ are 
$\eta$ = 1.88 and 1.86, respectively.
{\it Inset:} Steady-state vortex length $L_{\rm sat}$, also averaged over the oscillation 
period of the drive, as a function of the mutual friction parameter $\alpha$.
}
\label{f.spectrum}
\end{figure}

\section{Steady state spectrum}\label{s.results}

Our simulations are started from an initial state with an unperturbed vortex without Kelvin waves. 
The drive is then switched on and simulations are continued until a steady state is achieved. 
The results are summarized in Fig.~\ref{f.spectrum} where the steady-state Kelvin spectrum is plotted using 
different resolutions and different values of mutual friction. The spectrum is symmetric with respect 
to modes $m$ and $-m$, therefore only positive modes are plotted. At least our ``low'' and ``medium'' resolution 
calculations indicate that the absolute value for the Kelvin amplitude $|w_m|$ depends linearly on the drive 
amplitude. This was verified by increasing and decreasing the drive amplitude by a factor of five. 
In this respect our results differ from Ref.~[\onlinecite{VinenPRL2003}], where the 
Kelvin amplitude was observed to be independent of the drive amplitude. In our case dissipation 
is only due to mutual friction. Numerical dissipation (or noise) is quite small. For example, 
the kinetic energy is well conserved at $T=0$. The largest error comes from the calculation of the 
m.f. dissipation at scales that are close to the numerical resolution limit $m_{\rm res}$, where
the Kelvin-wave dispersion relation is no longer accurately modeled, as noted in 
Ref.~[\onlinecite{KS2005prl}]. A simple estimation gives that $\omega_m \propto 1-\cos(\pi m/m_{\rm res})$, 
which becomes flat when $m \sim m_{\rm res}$ and increases the timescale for modes near the 
resolution limit by a factor of $\pi^2/4$ when compared with Eq.~(\ref{e.kelvinfreq}) 
\cite{KozikPrivate}. Naturally, all the dissipation coming from scales smaller than our resolution
is also omitted in the calculations. Therefore the calculation of $P_{\rm mf}$ underestimates the real 
input power. At zero temperature in Ref.~[\onlinecite{VinenPRL2003}] the situation is more complicated 
in this respect since the determination of the input power is difficult. The amplitude of the 
oscillating superfluid velocity (which was used in Ref.~[\onlinecite{VinenPRL2003}] 
to drive the KWs) does not tell how much energy is really injected. In the pure Kelvin-wave 
cascade the absolute amplitude of the spectrum should still depend weakly on the amount of 
energy cascaded, $\epsilon$. 
Kozik-Svistunov's theory states that $n_m=|w_m|^2\propto \epsilon^{1/5}$, while L'vov-Nazarenko's 
formalism predicts that $n_m \propto \epsilon^{1/3}$. 

\begin{figure}[t]
\includegraphics[width=0.99\linewidth]{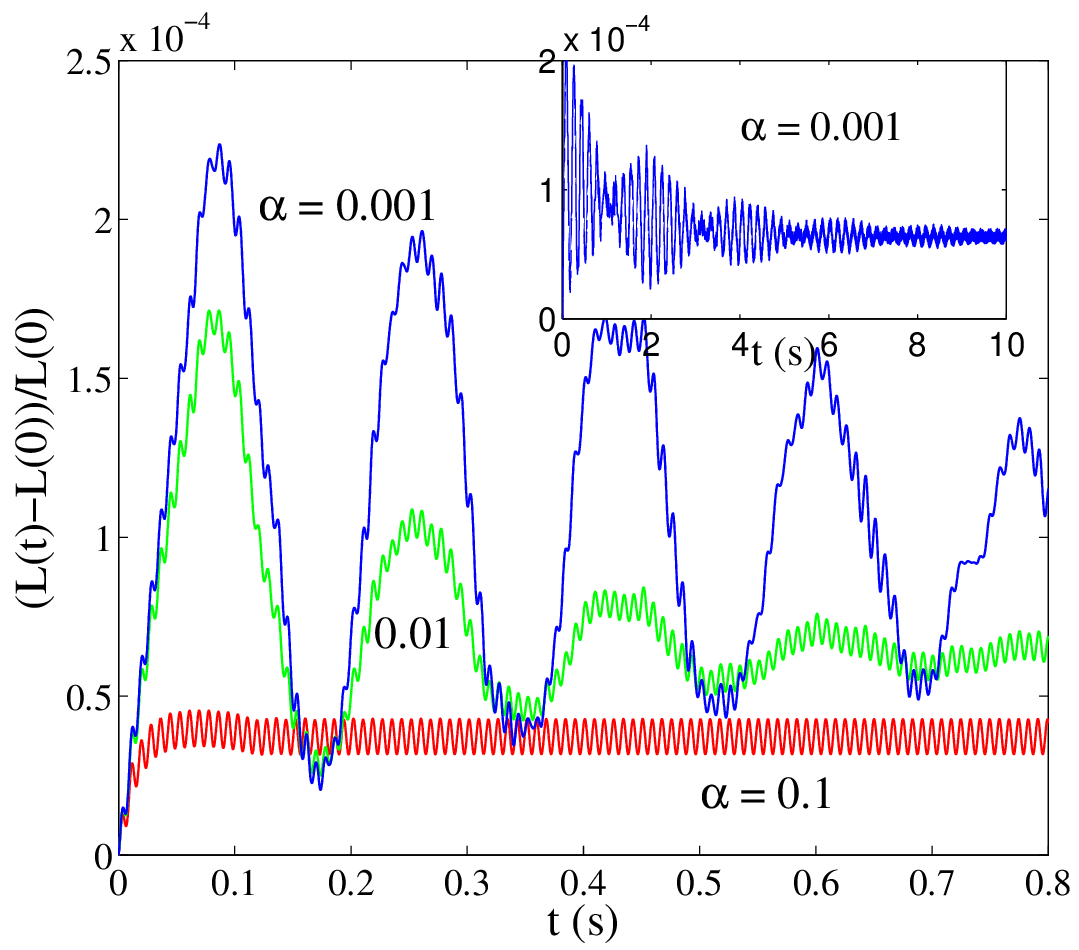}
\caption{(Color online) 
Transient response in vortex length $L(t)$ as a function of time when the drive 
is switched on. The drive amplitude is $x_{0}$ = 0.0005 mm and its frequency is the 
resonance frequency of mode $m$ = 10. 
The high frequency oscillations correspond to the frequency of the drive. 
The relaxation rate towards a steady state is mainly determined by mutual friction.
The inset illustrates the additional oscillations, which become visible when 
$\alpha$ = 0.001.
}
\label{f.length}
\end{figure}

As noted above, the timescales for modes near the resolution limit are longer than what
Eq.~(\ref{e.kelvinfreq}) states. One might expect that this explains why the spectrum 
bends up near $m_{\rm res}$. However, this is not the case since pinning three neighboring 
points (forcing these three points to move coherently), change the spectrum near $m_{\rm res}$,
while no effect is seen at lower $m$ values (Fig.~\ref{f.pinned}). Therefore 
the spectrum near the numerical resolution limit depends on the detailed structure 
of the pinning site. To consider length scales of order (and smaller) than the 
pinning site, one should take into account the geometry more precisely.  

The obtained spectrum with $\eta \approx 1.88$ is close but not equal to the theoretical 
prediction by L'vov and Nazarenko\cite{LN2010}, where $\eta = 11/6 \approx 1.833$. This 
could indicate that the energy cascade is non-local and dominated by different-scale 
interactions. It is more likely that the approximate match of the exponent is accidental 
and specific to our case with pinning sites. This is discussed in more detail in 
Sec.~\ref{s.boundary}, where we analyze the effect of the drive and the boundary conditions.

\section{Transients and relaxation towards the steady state}\label{s.transients}

\begin{figure}[t]
\includegraphics[width=0.99\linewidth]{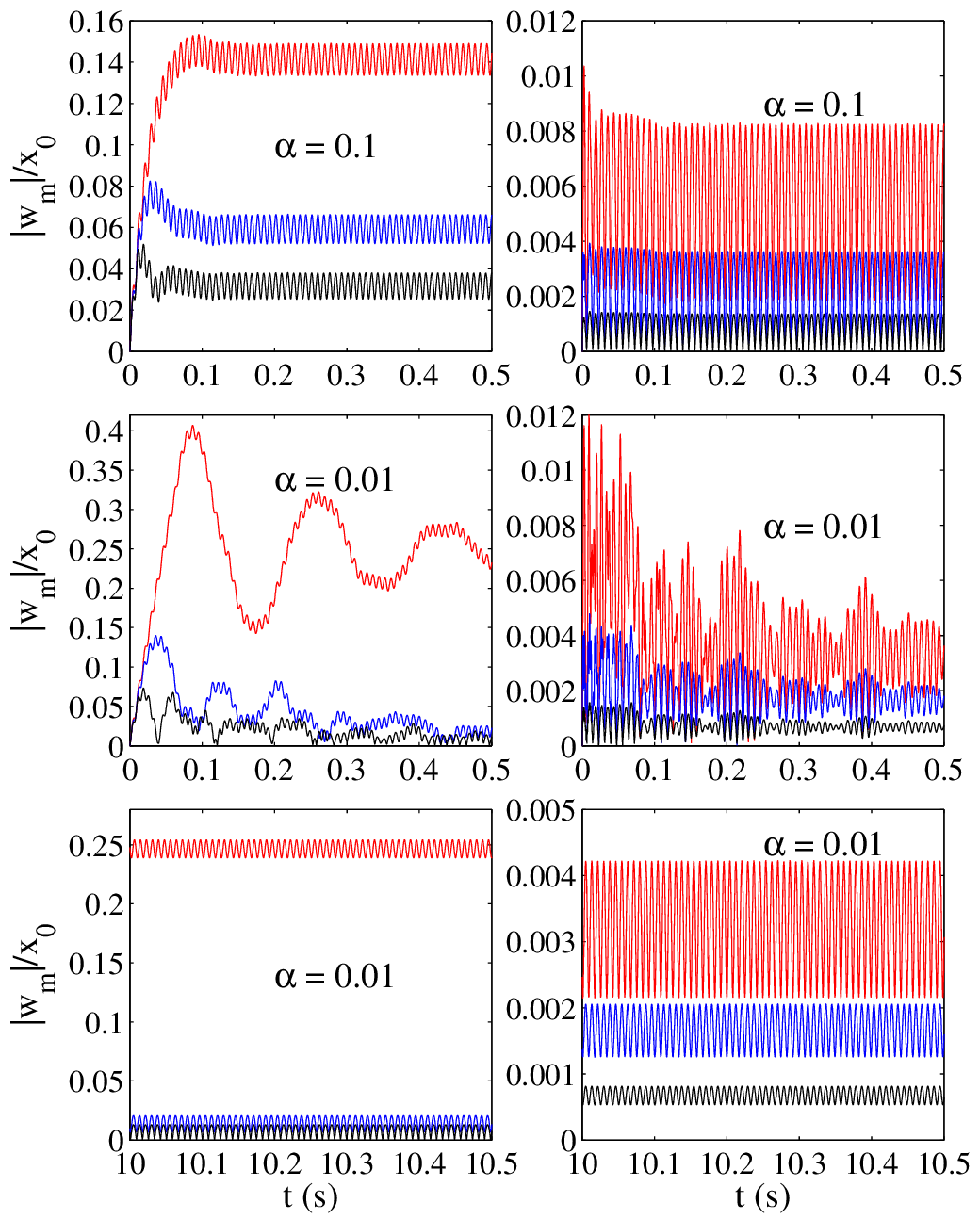}
\caption{(Color online)
Time development of the Kelvin amplitudes after switch-on of the drive. 
The drive amplitude is $x_{0}$ = 0.0005 mm and its frequency is near the resonance 
frequency of mode $m$ = 10. Left column of panels is for modes 10, 11, and 12 and 
right column for modes 20, 30, and 50. 
Note the different time intervals and scales.}
\label{f.amp}
\end{figure}

Due to the nature of the drive, the steady state is not a state where the vortex length and 
the amplitudes of the Kelvin waves are constant, but a state where they oscillate around 
their average values at a frequency determined by the drive frequency. These average values
for the Kelvin amplitudes are plotted in Fig.~\ref{f.spectrum}.  
The oscillations are illustrated in Figs.~\ref{f.length} and \ref{f.amp}.
The lower the temperature (smaller mutual friction parameter $\alpha$), the smaller is 
the oscillation amplitude (relative to the average value). Unfortunately, a smaller $\alpha$
also means a longer transient time to reach the steady state. The relaxation time towards a 
steady state is inversely proportional to the mutual friction parameter $\alpha$.
Therefore we are mostly limited to values $\alpha \gtrsim 0.001$. However, the simulations 
also indicate that the spectrum and the vortex length saturate below $\alpha=0.01$, which 
we interpret as the ``zero temperature limit''. Naturally, the value of $\alpha$, below which 
the spectrum does not change any more, should depend on the scales considered. If the drive 
frequency would be much higher (corresponding to shorter Kelvin waves), one could expect that 
a smaller value of $\alpha$ would be required to reach the ``zero temperature limit''. 

During the relaxation towards the steady state one may observe oscillations at frequencies
that are not related to the drive (Figs.~\ref{f.length} and \ref{f.amp}). 
These oscillations are visible in the vortex length but also in the mode amplitudes 
when the mode is close to the drive mode. Oscillations with period $\tau \approx 0.18$~s
are dominant for $\alpha$ = 0.01 and $\alpha$ = 0.001, but barely noticeable for
$\alpha$ = 0.1. For $\alpha$ = 0.001 one may notice additional beating  
with a very similar frequency (inset of Fig.~\ref{f.length}). 
It is not obvious whether these oscillations are related to Kelvin frequencies or not. 
The Kelvin period of mode $m$ = 3 is 0.17 s, but the neighboring modes are already quite far 
(periods related to modes 2 and 4 are 0.37 s and 0.096 s, respectively). Generally, 
it seems that more oscillations, with different frequencies, become visible as $\alpha$ 
decreases and their amplitudes increase. It is possible that in the limit of zero temperature 
these oscillations grow such that a steady state becomes impossible.

\section{Importance of the boundary conditions and the effect of the drive}\label{s.boundary}

\begin{figure}[t]
\includegraphics[width=0.99\linewidth]{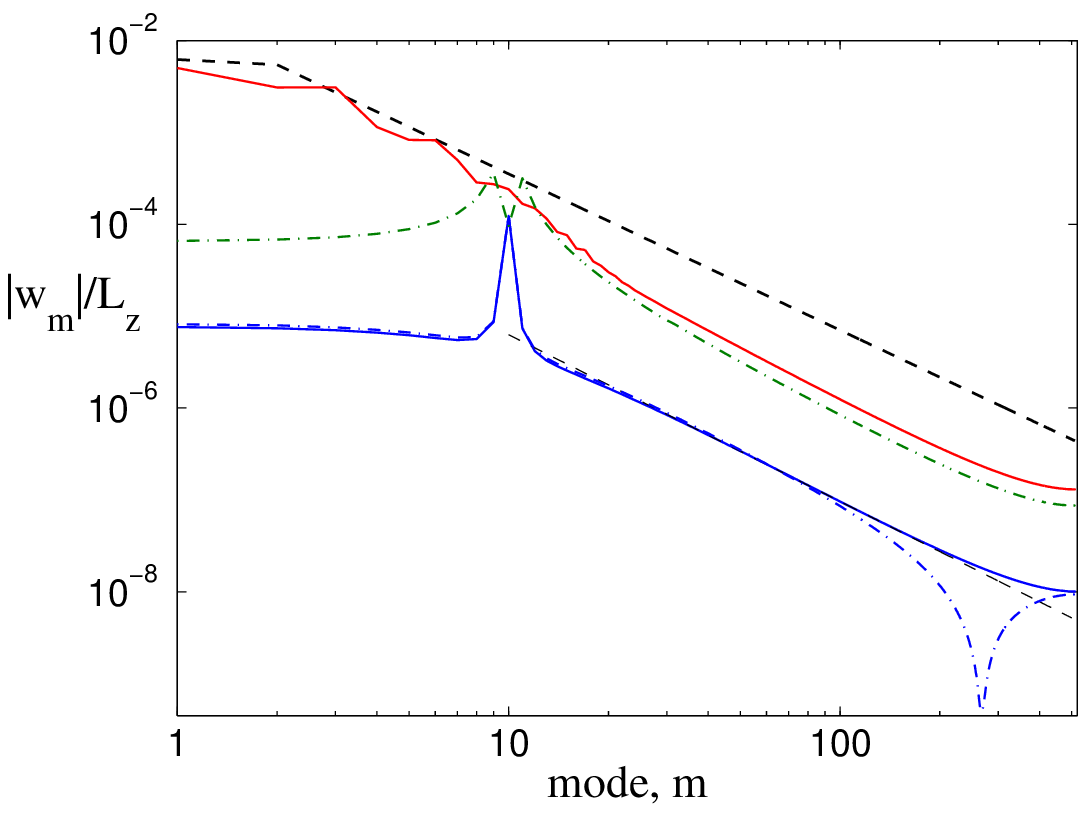}
\caption{(Color online)
Kelvin spectrum with pinned end points when $N$ = 1024 and $L_z$ = 1 mm. The lower (blue) solid curve is 
the steady state spectrum for $\alpha$ = 0.01 and $x_{0}=0.0005$ mm, presented already in Fig.~\ref{f.spectrum}. 
Similar lower (blue) dash-dotted curve is calculated using the same parameters, but driving not only one, 
but three neighboring points with the same amplitude. The upper (green) dash-dotted curve is
for a vortex that is driven by oscillating superfluid velocity of the form 
${\bf v}_s = v_0\cos(k_{10}z-\omega_{10}t)\hat{\bf x}$ with $v_0$ = 0.05 mm/s, also using $\alpha$ = 0.01. 
The upper solid (red) curve describes the time averaged (over 0.25 s) spectrum after a vortex, with initial 
Kozik-Svistunov spectrum ($\eta = 1.7$), has decayed 1.5 s with $\alpha$ = 0.001. The uppermost (black) dashed 
line illustrates the initial spectrum. The modes $m=\pm 1$ are modified to satisfy that $x=y=0$ at $z=0,L_z$. 
}
\label{f.pinned}
\end{figure}

The obtained spectrum is universal, independent of the drive, as long as the vortex 
end points are kept pinned. This is illustrated in Fig.~\ref{f.pinned}, where the same spectrum 
appears when we drive the vortex by oscillating superfluid velocity of the form 
${\bf v}_s = v_0\cos(k_mz-\omega_mt)\hat{\bf x}$, similar to one used in Ref.~[\onlinecite{VinenPRL2003}] 
(while keeping the end points fixed at $x=y=0$), 
or allow the initial (e.g. Kozik-Svistunov) spectrum to relax towards a straight vortex. 

\begin{figure}[t]
\includegraphics[width=0.99\linewidth]{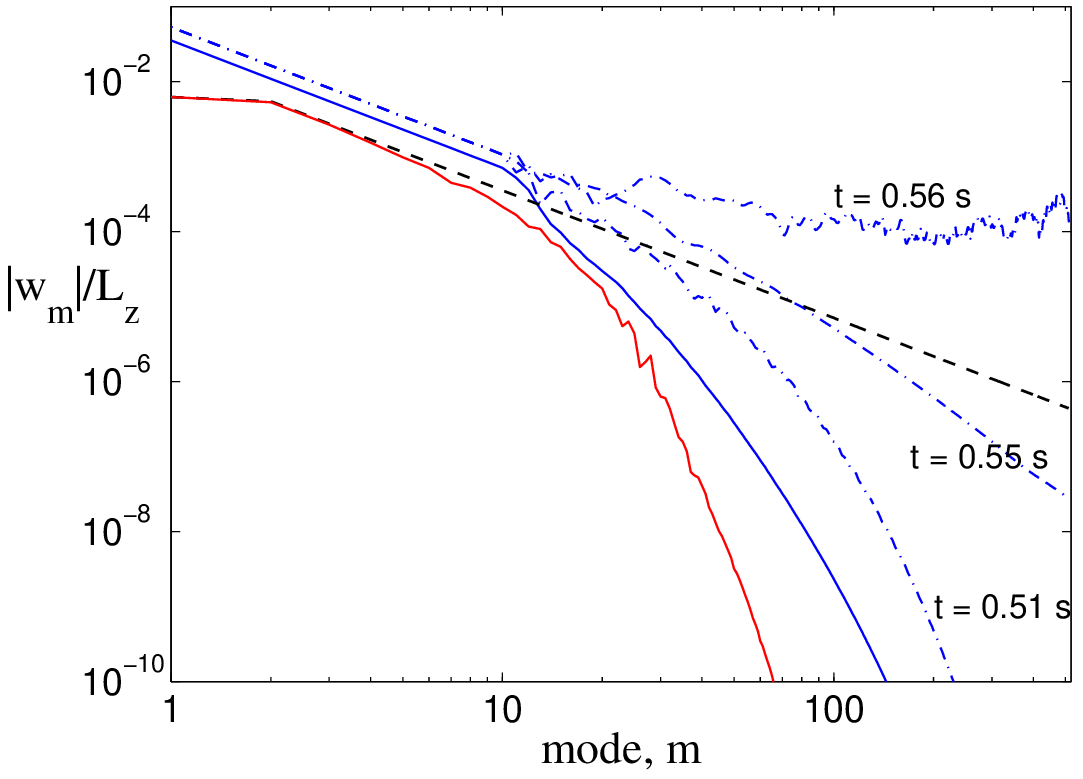}
\caption{(Color online)
Kelvin spectrum without pinning, but still assuming periodicity, when $N$ = 1024 and $L_z$ = 1 mm. 
The (black) dashed line is again the initial spectrum, with $\eta = 1.7$, for the decaying 
vortex and the lower (red) solid curve describes the time averaged (over 0.25 s) spectrum after the vortex 
has decayed 1.5 s with $\alpha$ = 0.001. The upper (blue) solid line describes the steady state spectrum 
when the vortex is driven in the $k$-space by periodically amplifying the lowest modes $|m| \leq 10$ using
a spectrum with $\eta = 1.7$ and amplitude $|w_{\pm 1}|/L_z = 0.035$ with $\alpha$ = 0.002. 
The (blue) dash-dotted curves illustrate how a small increase in the drive amplitude 
(to $|w_{\pm 1}|/L_z = 0.053$) causes that the configuration ``explodes'' such that soon the Fourier 
presentation fails.
}
\label{f.periodic}
\end{figure}

The situation is completely different if we omit pinning and allow the
end points to move freely (within the limitation of the periodicity). In this case a 
high-$k$ mutual friction cut-off appears and practically all the modes above the drive 
frequencies are exponentially damped. Few example spectra are shown in Fig.~\ref{f.periodic}.
We were not able to observe a steady state spectrum of type $|w_m|\propto m^{-\eta}$. 
Decreasing the mutual friction damping or increasing the drive typically lifts up the high-$k$ 
part of the spectrum, but at some point even a small change in the drive or in the damping changes 
the spectrum from the exponentially damped one to a case where the Fourier presentation fails. 
This is illustrated in Fig.~\ref{f.periodic}, where we drive the vortex in the $k$-space by 
amplifying the modes $|m|\leq 10$ (without changing the phase) periodically such that these modes 
have the Kozik-Svistunov spectrum with, $\eta = 1.7$. More numerical points with much wider range 
of scales could perhaps help to avoid the over-amplification of the highest modes (by increasing 
dissipation), but since the required work for a fixed time window scales like $N^3\ln{N}$ this 
becomes extremely time consuming to check.

The strong difference in spectra with and without pinning sites can qualitatively be understood 
in the following way. Pinned end points cause a strong boundary for the Kelvin waves. KWs cannot
travel freely but are typically reflected from the pinning site. This is similar to a case with 
plane boundary which prevents flow through it. Since we are considering small pinning sites we
are not forcing the vortex to terminate perpendicularly to the pinning site but fix it location. 
This implies a small cusp in vortex, which means that more small scales structures appear.

\begin{figure}[t]
\includegraphics[width=0.99\linewidth]{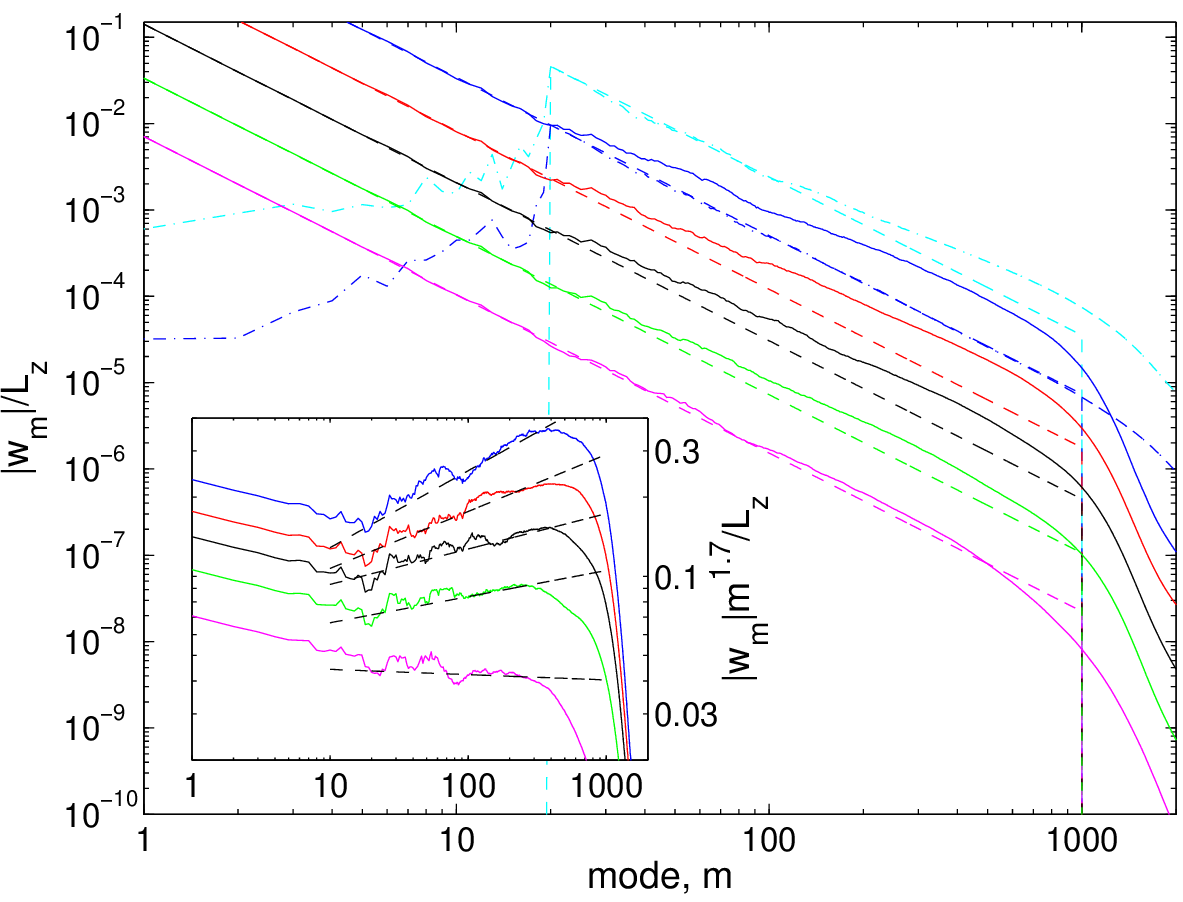}
\caption{(Color online)
Kelvin spectrum during the decay at $T=0$ when $N$ = 8192, $L_z$ = 1 mm, and the modes with $|m|>$ 1000 
are exponentially damped to mimic the energy sink at large $k$. The solid curves are time 
averaged (over $t$ = 0.015$\ldots$0.020 s) and $k$-smoothed spectra when the initial spectra 
(dashed lines) is $|w_m| = A|m|^{-\eta}$, with $\eta$ = 11/6. From bottom to top the initial amplitude, 
$A/L_z$, is 0.0707, 0.106, 0.141, 0.177, and 0.233. For visibility, the different curves have been 
shifted vertically by multiplying them by $1/10$, $1/\sqrt{10}$, $1$, $\sqrt{10}$, and $10$, 
respectively. The dash-dotted curves are for $A/L_z = $ 0.233 ($\times$ $10$) and 0.354 ($\times$ $10\sqrt{10}$) 
when initially the modes $|m| < 20$ are absent. For these runs the exponential damping of 
the highest modes is done less frequently since the energy flow is smaller and less damping is required. 
{\it Inset:} Compensated spectrum $k^{17/10}|w_k|$ together with fits (black dashed lines), which 
estimate that $\eta$ = 1.72, 1.60, 1.57, 1.48, 1.41, respectively. 
}
\label{f.kozik}
\end{figure}

Next we will shortly describe results from simulation that mimic the calculations done by Kozik and 
Svistunov \cite{KS2005prl,KS2010prbsub}. Here the mutual friction is omitted and the highest $k$-modes are 
periodically damped (exponentially) to mimic the energy sink. Pinning is omitted and periodic boundary
conditions are used. The initial spectrum is L'vov-Nazarenko spectrum with $\eta$ = 11/6 and we follow 
the time development using different amplitudes for the initial spectrum. Figure \ref{f.kozik} summarizes 
these results. 

As one can see, the spectrum depends on the amplitude of the initial spectrum, as predicted 
by Sonin recently\cite{Sonin2012}. The larger the initial amplitude, the smaller $\eta$ is observed. 
The smallest value for $\eta$ = 1.41 is already so small that the vortex length would diverge without a 
cutoff near the core size. It is also a bit smaller than the lower limit predicted theoretically\cite{Sonin2012}. 
These values for $\eta$, for a given amplitude, are only approximate since the spectrum 
still slowly changes with time. For example, the transition region from the low $k$-region (having 
the initial spectrum) to the high $k$-region with different $\eta$ is quite well determined by equating
the iteration time with the Kelvin period. The exponent $\eta$ in the high $k$-region still has 
a weak time dependence. To get rid off all the transients one should continue iterations much 
further. However, without a drive the true steady state is difficult to reach. In any case, 
these simulations illustrate that the Kelvin spectrum is not unique, but depends on the amplitude. 
Theoretical predictions by L'vov-Nazarenko and Kozik-Svistunov were done in the zero amplitude 
limit, which might be difficult to realize experimentally. 

The dash-dotted curves in Fig.~\ref{f.kozik}, where the modes $|m| < 20$ are initially absent, also 
illustrate that spectrum is not simply determined by the amplitude but is also affected by the lowest 
modes. This would indicate that the cascade is non-local (and affected by the large scale curvature), 
supporting in this respect the theory by L'vov and Nazarenko. In any case, the spectrum is not 
as universal as one would expect from papers [\onlinecite{KS2005prl,KS2010prbsub}].

\section{Absorption from mutual friction}\label{s.Pmf}

At high temperatures vortex motion is typically laminar and dissipation is only due 
to mutual friction. At low temperatures turbulence becomes more common and dissipation
increases. In the zero temperature limit all the energy is cascaded to scales of order the 
vortex core size and is dissipated by phonon emission ($^4$He). But at finite temperature, 
before the zero temperature limit is reached, the small scale Kelvin waves can greatly 
increase the m.f. dissipation when compared to the laminar case. This was seen
in the simulations with oscillating pinning points (Secs. \ref{s.results} and \ref{s.transients}) 
where the dissipated power was strongly affected by the resolution used in the simulations.  
In order to see how fast the power dissipated increases with the number of Kelvin modes we 
integrated Eq.~(\ref{e.Pmf}) by populating a vortex of length $L_z$ = 1 mm with Kelvin 
waves, using up to 131072 discretization points on a vortex. Periodic b.c. are assumed. 
Note that here we do not pin (nor drive) the vortex end points. We just populate the 
vortex with a a spectrum $|w_m| \propto m^{-\eta}$ and include Kelvin waves up to some 
maximum mode $N_{\rm KW}$ ($N_{\rm KW} \ll m_{\rm res}$), which corresponds to a cutoff 
scale $\xi_{\rm cutoff} = L_z/N_{\rm KW}$. The phase of each mode is set randomly and 
the amplitude for modes $m=\pm{1}$ are $|w_{\pm{1}}|/L_z = 0.01/\sqrt{2}$. The positive
and negative modes are chosen to have equal amplitude, i.e. $|w_{m}|=|w_{-m}|$. This
choice is not important, e.g. by choosing a random distribution between negative and 
positive modes changes the results only by few percent.  

\begin{figure}[t]
\includegraphics[width=0.99\linewidth]{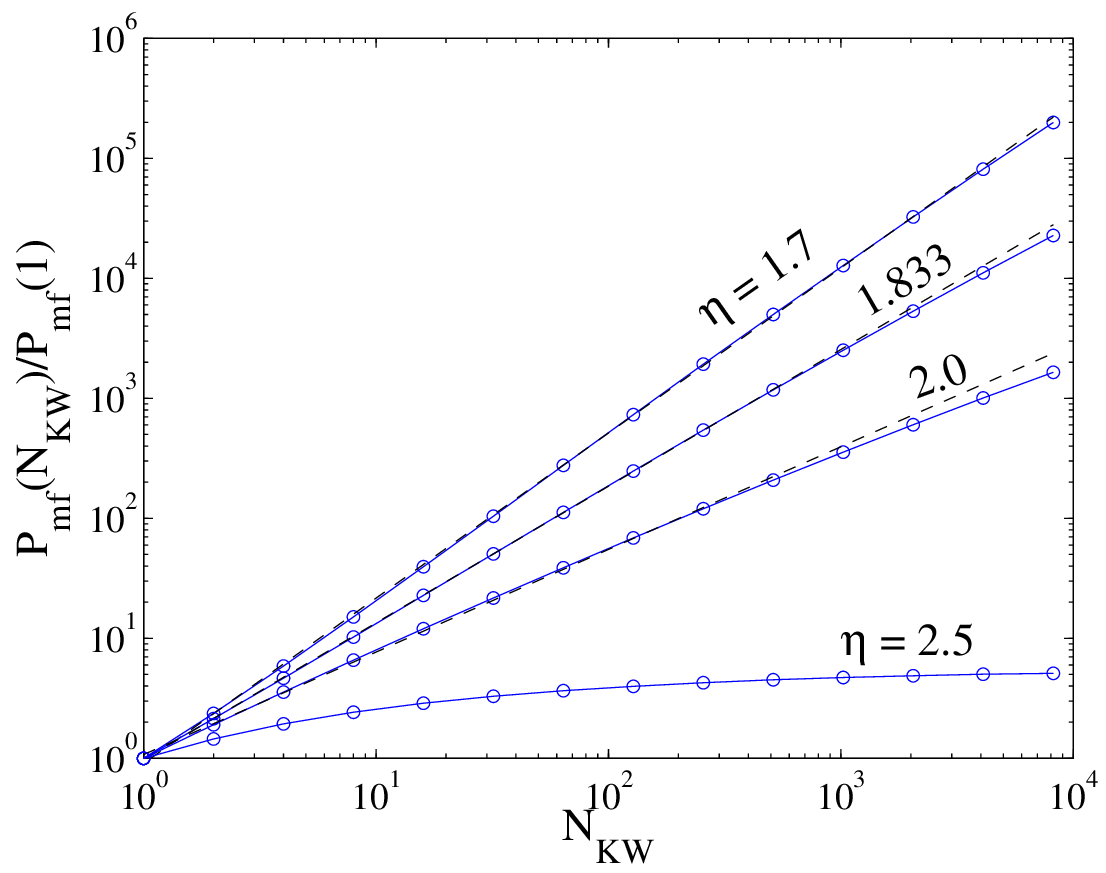}
\caption{(Color online) 
Power dissipated by mutual friction as a function of the largest Kelvin mode
$N_{\rm KW}$ for different spectra $|w_m| \propto m^{-\eta}$. The dashed lines 
show the best fits according to Eq.~(\ref{e.pfit}) when the four highest 
data points (blue circles) in each set are omitted from the fit.    
}
\label{f.Pmf}
\end{figure}

We observed no saturation in the dissipated power for $\eta \lesssim 2$, at least 
for Kelvin spectra reaching modes up to $N_{\rm KW}$ $\sim$ 8000, which corresponds 
to a cutoff scale of order $\xi_{\rm cutoff} \sim 0.1$ $\mu$m. 
For $\eta = 2.5$ the dissipated power might saturate as seen in Fig.~\ref{f.Pmf}, where 
the power dissipated is plotted as a function of $N_{\rm KW}$ for a few different KW-spectra. 
A simple estimate, using the local induction approximation (LIA) where 
${\bf v}_s = (\kappa/4\pi) \ln(R/a) {\bf s}'\times{\bf s}''$, 
gives that the power converges, in the limit of $N_{\rm KW}\rightarrow \infty$, 
only when $\eta > 5/2$.

The dissipated power depends linearly on the m.f. parameter $\alpha$ and, for small 
amplitudes, quadratically on the amplitude of the Kelvin waves. For a wide range of 
scales one can see from Fig.~\ref{f.Pmf} that for $\eta\lesssim 2$ the power dissipated 
due to m.f. is approximately given by
\begin{equation}\label{e.pfit}
P_{\rm mf} = \alpha C \xi_{\rm cutoff}^{-\beta},
\end{equation}
where $C$ depends at least on the amplitude of the KWs. We determined that 
$\beta \approx$ 1.38 for the Kozik-Svistunov \cite{KS2004prl} spectrum ($\eta$ = 17/10) 
and $\beta \approx$ 1.14 for the L'vov-Nazarenko \cite{LN2010} spectrum ($\eta$ = 11/6). 
The above equation implies that, if $P_{\rm mf}$ is kept fixed by the drive and 
the spectrum is assumed to change only by a different (sharp) cutoff scale $\xi_{\rm cutoff}$, any 
change in $\alpha$ from $\alpha_1$ to $\alpha_2$ results in a corresponding change in $\xi_{\rm cutoff}$ 
from $\xi_{\rm cutoff,1}$ to $\xi_{\rm cutoff,2}$ which are related as 
\begin{equation}
\frac{\xi_{\rm cutoff,2}}{\xi_{\rm cutoff,1}} = \left(\frac{\alpha_2}{\alpha_1}\right)^{1/\beta}.
\end{equation}
Since the exponent $1/\beta$ is close to 1, any change in $\alpha$ will cause 
a similar change in the cutoff scale.

If one could obtain reasonable estimates for the cutoff scale and for the amplitudes 
of the lowest modes, one could try to estimate $\eta$ by measuring the power dissipation.
In practice this might be difficult. 

In summary, our calculations for power dissipation indicate that the mutual friction can
dissipate lots of energy. This is especially true for the case of $^4$He where the core 
size is atomic and one can estimate that $N_{\rm KW} \sim \ell/a \sim 10^6$, where 
$\ell \sim 0.1$ mm is the average vortex spacing.

\section{Discussion}

The calculated spectrum in Sec. \ref{s.results} should correspond to an experimental situation 
where the pinning site is pointlike. 
An interesting point of these simulations is that there appears no m.f. dissipation dependent 
cutoff at high k-values, even if we use a relatively high value 0.1 for $\alpha$. This is 
also caused by the spatially peaked drive and the fact that we are fixing the oscillation 
amplitude and not the power. By improving the numerical resolution, the power dissipated by 
mutual friction (which estimates the input power) increases quite 
rapidly with the number of Kelvin modes present, that is determined by our resolution. 
One should remember that the number of modes is limited by a physical cutoff 
determined by the vortex core size. At smaller scales, at least, the type of dissipation 
should change from m.f. dissipation to something else (e.g. phonons
in case of $^4$He). Unfortunately, the vortex filament formulation does not allow us to 
approach this limit.

The Kelvin-wave cascade is expected to be a rather inefficient method for transferring energy 
to smaller scales and thus one might expect a bottleneck to appear in the energy transfer rate 
if the input power is too large \cite{LvovPRB2007}. Unfortunately, the ideal case to observe 
the bottleneck effect, low temperature and high drive, is more demanding numerically and typically 
results in a failure to present the vortex configuration as a Fourier series. Perhaps this failure 
is an indication of the breakdown of the cascade and the accumulation of Kelvin waves, owing to 
the appearance of the bottleneck. 

Numerical calculation of the energy flux through different length scales is a challenging
problem since the flux due to Kelvin-wave cascade is extremely weak.  
Without a proper estimate for the energy flux it is difficult to quantify which part (or
fraction) of the spectrum is generated by the Kelvin-wave cascade and what is the effect 
of the drive or due to pinning. As illustrated above, pinning tends to give a 
particular spectrum, independent of the drive. Therefore, the spectrum cannot be much 
affected by drive, but can be dominated by the effects due to pinning. 
By shaking the end points one forces an additional small U-shaped contour on the vortex. 
For example, a parabolic shape would result in a spectrum with $|w_m|\propto m^{-2}$, when 
interpreted as a sum of Kelvin waves. However, since same spectrum appears for a decaying
vortex, where this U-shaped contour is absent, the spectrum must be affected by a small cusp 
appearing at the pinning point and that causes a discontinuity in the derivative of the vortex 
configuration ${\bf s}(\xi)$. This discontinuity also results in a spectrum of type $|w_m|\propto m^{-2}$. 
However, sharp cusps in the vortex configuration are unavoidable at low temperatures with vanishing 
mutual friction. They often result from reconnections that are expected to drive the Kelvin cascade.

\section{Conclusions}

We have determined the Kelvin spectrum in a realistic, experimentally feasible situation 
by driving the Kelvin waves by shaking the vortex end points and damping the high 
$k$-values by mutual friction. Therefore our results provide an improvement over previous 
work. When the vortex end points are kept pinned the obtained spectrum at scales larger 
than the pinning site is quite robust: independent of the drive and quite insensitive to 
temperature. Without pinning the spectrum is not unique. For example, the Kelvin spectrum 
for a vortex that is allowed to relax on its own at $T = 0$ produces a spectrum that 
depends on the absolute amplitude of the initial spectrum. Therefore, it is expected that
in real systems the Kelvin spectrum will depend on external conditions.

\begin{acknowledgments}
I am extremely grateful to M. Krusius and V.~B. Eltsov for their comments and improvement for 
the paper.  I also thank N. Hietala, J. Karim\"aki, and E. Kozik for their comments. This work is 
supported by the Academy of Finland (grant 218211) 
and EU 7th Framework Programme (FP7/2007-2013, grant 228464 Microkelvin).
\end{acknowledgments}

\end{document}